# A scalable method for molecular network reconstruction identifies properties of targets and mutations in acute myeloid leukemia

Edison Ong[1], Anthony Szedlak[2], Yunyi Kang[3], Peyton Smith[1], Nicholas Smith[1], Madison McBride[3], Darren Finlay[3], Kristiina Vuori[3], James Mason[4], Edward D. Ball[5], Carlo Piermarocchi[2]*, Giovanni Paternostro[3]*

[1]Salgomed Inc., Del Mar, CA 92014
[2]Department of Physics and Astronomy, Michigan State University, East Lansing, MI 48824
[3]Sanford-Burnham Medical Research Institute, La Jolla, CA 92037
[4]Scripps Health, San Diego, CA 92121
[5]University of California, San Diego Moores Cancer Center and Department of Medicine, La Jolla, CA 92093

* Corresponding authors: piermaro@msu.edu and giovanni@sanfordburnham.org

**ABSTRACT**

A key aim of systems biology is the reconstruction of molecular networks, however we do not yet have networks that integrate information from all datasets available for a particular clinical condition. This is in part due to the limited scalability, in terms of required computational time and power, of existing algorithms. Network reconstruction methods should also be scalable in the sense of allowing scientists from different backgrounds to efficiently integrate additional data.

We present a network model of acute myeloid leukemia (AML). In the current version (AML 2.1) we have used gene expression data (both microarray and RNA-seq) from five different studies comprising a total of 771 AML samples and a protein-protein interactions dataset. Our scalable network reconstruction method is in part based on the well-known property of gene expression correlation among interacting molecules. The difficulty of distinguishing between direct and indirect interactions is addressed optimizing the coefficient of variation of gene expression, using a validated gold standard dataset of direct interactions. Computational time is much reduced compared to other network reconstruction methods. A key feature is the study of the reproducibility of interactions found in independent clinical datasets.

An analysis of the most significant clusters, and of the network properties (intraset efficiency, degree, betweenness centrality and PageRank) of common AML mutations demonstrated the biological significance of the network. A statistical analysis of the response of blast cells from eleven AML patients to a library of kinase inhibitors provided an experimental validation of the network. A combination of network and experimental data identified CDK1, CDK2, CDK4 and CDK6 and other kinases as potential therapeutic targets in AML.

The most updated version of the AML network can be found at www.leukemianetworks.org

## INTRODUCTION

The knowledge of genomic changes and of other "omic" alterations in patients with acute myeloid leukemia (AML) has increased significantly over the last decade (1, 2). This has not, however, resulted in the development of new effective therapies and AML still has an unfavorable prognosis for most patients (1). Robert Weinberg, one of pioneers of the reductionist molecular approach to cancer research (3), recently suggested that new data-rich approaches are needed to address the complexity and heterogeneity of cancer. He, however, also pointed out that systems biology has not yet led to major advances in the understanding and treatment of malignant neoplastic disease.

A key aim of systems biology is the reconstruction of informative molecular networks and it is becoming clear that only cell-specific and disease-specific networks are potentially able to benefit medical practice (4 , 5). These networks could be used, for example, to obtain actionable information from the complex and often unique mutational cancer profiles that sequencing data provide (2).

More than one million gene expression datasets are available in public repositories (6) and biology is clearly ready for the Big Data computational approaches that are increasingly used in other fields of science and technology (7, 8). We do not yet, however, have disease-specific networks that integrate information from most available datasets. This is in part due to the limited scalability, in terms of required computational time and power, of existing algorithms.

It is also becoming clear that integrating the growing number of datasets and the increasing amount of knowledge for a particular pathology is not a realistic task for individual research groups or even companies. Network reconstruction methods should therefore also be scalable in another sense: they should allow scientists from different groups and background to efficiently integrate additional data and progressively increase the network accuracy.

In the version of the AML network we describe here (version 2.1) we have used gene expression data (both microarray and RNA-seq) from five different studies (9-13) comprising a total of 771 AML samples. We also integrate a human protein-protein interactions dataset (14).

The method we present is in part based on the well-known property of gene expression correlation among interacting molecules in biological networks (15). A potential problem is the difficulty of distinguishing between direct and indirect interactions. We suggest a solution based on the optimization of a statistical property, the coefficient of variation, using a validated "gold standard" dataset of direct interactions. We show that computational time is much reduced compared to other network reconstruction methods and that adding new datasets is especially easy because most computations already performed do not need to be repeated. We also suggest statistical measures that can provide the optimal correlation coefficient cutoff for the selection of significant interactions. A key feature of the method is "overlap analysis", which is based on the study of reproducibility of interactions found in two or more independent clinical datasets.

An analysis of the most important clusters, and of the network properties of receptors and of common AML mutations demonstrates the biological significance of the network. The network properties of kinases are consistent with a statistical analysis of the experimental response of AML primary patient cells to a kinase inhibitor library and the two measures can be combined to identify potential targets for therapeutic interventions.

## MATERIALS and METHODS

### AML Molecular Network Reconstruction

Figure 1 shows the main modules of the procedure used to generate the AML molecular network (AML version 2.1) (more details of the integrations of the modules are shown in Figure 1S). The method consists of five major sections, indicated by different colors as step I to V: (I) Data Processing, (II) Optimization and Method Selection, (III) Transcription Factor-Gene (TFG) Subnetwork Reconstruction, (IV) Protein-Protein Interaction (PPI) Subnetwork Reconstruction and (V) Full AML Molecular Network. The TFG subnetwork is composed of transcription factors (TFs) and their targets, while the PPI subnetwork contains interactions between proteins. The TFG and PPI subnetworks were constructed using the five individual expression profiles described in the next section and later integrated to form the full AML molecular network. The resulting network is a partially directed network. The most updated version of the AML network can be found at www.leukemianetworks.org

### Data Processing

The data processing step is indicated as step I (with yellow coloration) in Figure 1. Three AML microarray datasets (9-11) were downloaded from the Gene Expression Omnibus (GEO) database as raw data (CEL files) and processed using the *threestep* function of the affyPLM R package with default settings. Median gene expression was used whenever multiple probes were mapped to a single gene. Two AML RNA-seq datasets (12, 13) were downloaded from GEO and from The Cancer Genome Atlas (TCGA, downloaded from the TCGA Data Portal on February 11, 2014) databases in the RPKM (reads per kilobase per million) format. Genes with RPKM below 1 were excluded and a logarithmic transformation was then used. A total of five different expression datasets were used. The five expression datasets were used to create five TFG and five PPI lists of interactions as described below and were later integrated in the TFG and PPI subnetworks.

### Optimization and Method Selection

Optimization and method selection is indicated as step II in orange in Figure 1. A key assumption of the optimization and method selection procedure is that the number of experimentally validated interactions (expressed as a ratio of the total number of interactions inferred) should be maximized. We used this assumption to establish criteria for noise filtering, cutoffs, and consistency checks for networks obtained from different datasets and different inference methods. For this purpose, we used the TRANSFAC database (version 2013.2) (16) as a "gold standard" of experimentally validated interactions and focused on a subset of interactions involving 90 transcription factors randomly selected among those present in the five AML expression datasets (9-13) and their targets.

A total of 2486 interactions and 1273 targets were found for these 90 transcription factors (see Table 1S) in TRANSFAC. These 2486 TRANSFAC Interactions (TIs, they can also be considered as "True Interactions") are experimentally validated in humans (17) and we used them as a "gold standard" in Optimization and Method Selection and TFG Subnetwork Reconstruction. All genes present in these TIs were selected from the expression datasets and five test expression profiles were created. These test expression profiles were used for the coefficient of variation (CV) optimization and for the methods comparison.

The CV is the standard deviation divided by the average of the expression of a gene across all experiments in one dataset. Eliminating genes with low CV removes noise, filtering out genes with low variation in their expression. Figure 2 (panel B) shows an example of the type of data that were removed. To determine an optimal value for the CV cutoff, we maximized the number of "gold standard" TIs found within the top 100 most significant interactions found by four well-known network inference methods. The four methods are ARACNE (18, 19), TIGRESS (20), GENIE3 (21) and Pearson Correlation (implemented in Python Scipy) (15). Figure 2 (panels C and D) shows an example of this optimization procedure.

Interactions inferred by these four methods were compared and their TI content (expressed as a ratio) is reported in Table 2S. The number of TIs found by using one of the other three methods after the Pearson Correlation approach is also reported in Table 2S. Note also that transcription factors usually promote expression of their targets, so expression values of proteins with this functional relationship should be positively correlated. We did, in fact, find an enrichment of the number of TIs (that is, validated transcription factors and their targets) for higher Pearson Correlation values (see Figure 3).

A run-time analysis was done for the four methods and, besides correlation, the other three methods were run using published software (19-21). The run-time analysis was done on a system with a AMD FX-4130 quad-core 3.80GHz processor and 32GB RAM. For this analysis we used the expression results from Eppert et al (9) as the testing dataset. Two variables, number of interactions and number of experiments, were used to measure how the size of the dataset affects the run-time (see Figure 2S). The interaction number was measured at a constant ratio, where the number of targets was 10 times more than the number of transcription factors (TFs), which is the approximate ratio in human cells (22).

Based on these analyses, the Pearson Correlation method was selected for the TFG and PPI subnetworks reconstruction.

**TFG Subnetwork Reconstruction**

The steps for reconstruction of the TFG subnetwork are shown as step III in blue in Figure 1. Optimized CV cutoffs obtained as explained in the previous section were used to create five expression profiles. Animal TFDB (23) and KEGG (24) were used to obtain the list of known transcription factors (1595 TFs) (Table 1S). Five TFG lists of interactions were inferred from the five expression profiles using Pearson's correlation and were ranked based on their correlation coefficients.

We then partitioned all the interactions in bins of equal size $N_B$ corresponding to different intervals of the correlation coefficient. To estimate the significance of the interactions in a given correlation interval we considered as a control the case in which the $N_B$ interactions are chosen randomly from a pool of $N$ possible interactions, and we calculated the probability of finding a certain number of TIs (the previously described "gold standard" TRANSFAC Interactions). Assuming a statistical model with replacement, we can estimate the probability that a TI is found by chance as $q = N_T/N$, where $N_T$ is the number of TIs. The probability that the bin finds $k$ true interactions by chance is then $P(k) = q^k(1-q)^{N_B-k}\binom{N_B}{k}$, which can be approximated by a Poisson distribution $P(k) = \frac{1}{k!}(\lambda)^k e^{-\lambda}$ where $\lambda = N_B q$ since $N_B \gg 1$ and $q \ll 1$.

The Poisson statistics was then used to determine the cutoff for the correlation coefficient. Figure 3A shows how the procedure was used for one of the datasets. The interactions were ranked according to the correlation coefficient and subdivided in bins containing the same number of interactions. Bins are ordered in Figure 3A according to the correlation coefficient values of the interactions they include decreasing from left to right. We then counted the number of TIs in each bin (this number is shown in the vertical axis in Figure 3A). In these bins we have 6000 interactions and in the random case we would expect an average of $\lambda = 2$ TIs. Using the Poisson distribution, the probability of having more than 3 TIs in a bin is less than 0.1 (Figure 3S shows the details of this distribution). To determine the cutoff we simply examine each bin from left to right and stop before the first bin with 3 or less TIs. The vertical red line in Figure 3A shows this cutoff.

As shown in Table 3S, the bin size and significance level for the cutoff were optimized by examining the effects on the overlap analysis, which will be described in detail in the next section. Briefly, we identified the parameters that maximized both the number of interactions shared by different datasets (we refer to these shared interactions as overlaps) and the number of TIs in these sets of overlapping interactions.

As in our procedure, it is common for the initial steps of a search considering multiple factors to select broad sets. An example of this is the implementation of web search by modern search engines (25). The rationale is that if a factor is not the only predictor, implementing it with a stringent cutoff eliminates many potentially useful elements (in our case interactions). This is shown by the optimization procedure we have mentioned in the previous paragraph and also by the fact that, as will be seen, even the set of interactions shared by all datasets includes interactions with relatively low correlation values. Also note that our procedure allows us to choose a different value of correlation as cutoff for each dataset and we can therefore take into account the possible variations in quality of the datasets. As table 3S shows we were indeed able to select more interactions on average from the RNA-seq datasets, which are considered to provide higher quality gene expression data compared to microarrays (26).

*Reproducibility analysis (overlap)*

A key aspect of our procedure takes into account the reproducibility of interaction identification from different clinical datasets. The five selected lists of TFG interactions were combined into a unique subnetwork. The interactions were separated into five groups: those found from one dataset only, those found from two datasets, and so on, up to those found in all five datasets. The shared interactions are from now on referred to as "overlaps" for brevity and the groups containing interactions shared by the same number of datasets are referred to as overlap groups. These overlap groups are shown in Table 4S.

To estimate the significance of the overlaps obtained from the five datasets we considered as control the case in which each of the five datasets yields $M$ random interactions chosen from a pool of $N > M$ possible interactions. Given one interaction, the probability that one dataset randomly finds it is $p = M/N$, and the probability that only one dataset out of five finds it is $P_1 = p(1-p)^4 \binom{5}{1}$. In the general case of $k$ datasets, this probability is $P_1 = p(1-p)^{k-1} \binom{k}{1}$, which gives an estimated number of interactions in the first group $N_1 = NP_1 \sim kM$. Similarly, the probability that one interaction is found by $j$ out of $k$ datasets is $P_j = p^j(1-p)^{k-j} \binom{k}{j}$ and the number of interactions in group $j$ is $N_j = NP_j \sim N \binom{k}{j} \left(\frac{M}{N}\right)^j$. This expression can be generalized to the case of datasets leading to networks with a different number of interactions $M_1, M_2, \dots, M_k$ as $N_j = \frac{1}{N^{j-1} j!} \sum_{\alpha_1 \neq \alpha_2 \neq \dots \neq \alpha_j} M_{\alpha_1} M_{\alpha_2} \dots M_{\alpha_j}$, where each index $\alpha_i$ runs from 1 to $k$.

The TFG subnetwork was further compared to 100 randomly generated subnetworks and the results were very similar to the estimates obtained as described in the previous paragraph (Tables 4A and 4B). The randomly generated subnetworks had the same number of interactions for each dataset as the TFG.

Because of our emphasis on reproducibility, only interactions that appeared in two or more datasets were selected for inclusion in AML 2.1. Among reproducible interactions, all interactions in a group (meaning the four groups composed of interactions found from two to five times) were included in the network if the number of interactions in the random simulation was less than 1% of the number of interactions found in the group. This implies that the number of interactions identified more than once only by random chance was negligible. If, however, the number of random interactions was higher than the 1% threshold, then the average correlation coefficient distribution was calculated for the random TFG subnetwork. Interactions with correlation coefficients below those corresponding to a p-value of 0.005 were removed. This method was used to select 6117 out of 17,817 interactions obtained from only 2 datasets. The average rank by correlation coefficient value was used to order the interactions and the top 6117 were selected. This set of interactions had a higher TI ratio (more than 3.4 fold higher) compared to those eliminated in the same group with overlap 2 (Table 4S).

We also analyzed 7 more AML microarray datasets (GSE15434, 24006, 33223, 34860, 21261, 6891, 22845) which will be included in future versions of the network, and measured the number of interactions from the overlap 2 group that were replicated again when additional studies were added. This analysis, shown in depth in the methodological results section, can assist the choice of which interactions to select when dataset numbers increase.

All interactions found by 3 to 5 expression profiles were kept in the TFG subnetwork. The edges between TF and their targets were considered as directed.

**PPI Subnetwork Reconstruction**

These procedures are shown as step IV in green in Figure 1. Protein-protein interactions were downloaded from HIPPIE (14), which is a database of experimentally validated protein-protein interactions. The Pearson correlation coefficients of all HIPPIE interactions were calculated using the 5 gene expression profiles (9-13) to infer the PPI ranked lists of interactions. The same correlation coefficient cutoffs obtained for the 5 TFG-ranked lists of interactions were also used as the cutoffs for the five PPI-ranked lists of interactions. Figure 3B-C shows that an enrichment for validated interactions with higher correlation coefficient values is also found for protein-protein interactions (see Table 5S), using the HIPPIE database as an independently validated list. These findings support our plans to extend to PPI the optimization approach used for TFG, in a future version of the network.

The integration of the 5 individual PPI datasets was done by using a similar overlap analysis as that for the TFG subnetwork. No group had a number of overlapping interactions in the random simulations that were higher than 1% of the total found for that group and therefore all interactions found in 2 to 5 expression profiles were selected for the PPI subnetwork. Interaction directionality was introduced, where possible, using the Phosphosite and Phosphopoint kinase/target databases (27, 28).

**Integration and Analysis of combined AML Network**

By combining the TFG and PPI subnetworks, a partially directed AML molecular network was generated and named AML 2.1.

Networks were visualized and analyzed using Cytoscape (29). The AML network was analyzed using the Cytoscape clustering app, ClustViz, which uses the graph theoretic method MCODE (30). The clusters were then analyzed with BiNGO (31) for enrichment analysis of gene ontology. Genes were also ranked based on their degree and other network measures and characterized by 4 functional categories: TFs, metabolic genes, kinases and receptors. Additionally, a metabolic pathway enrichment analysis was done using RECON2 (32). A comparison was made between normal human hematopoietic cells and AML cells, using ClusViz clusters and RECON2 pathways. The Fisher's exact test (from SciPy) was used for this comparison with the gene expression data from the Macrae et al. dataset (12), which also contained normal control data.

A mutation subnetwork was created using the commonly mutated AML genes (2) that were found in the AML 2.1 network (21 out of 26 were found) and their corresponding first neighbors.

## Definition and Analysis of Network Measures

The network measures (25) were calculated with the software provided by Rubinov and Sporns (33). The definition of *clustering coefficient*, *transitivity*, *assortativity* and *betweenness centrality* in Table 5 are from Rubinov and Sporns (33). Several network measures were confirmed using NetworkX (34). The following measures were calculated using NumPy and SciPy. For *efficiency* we introduced a new quantity, the *intraset efficiency* that generalizes the concept of *global efficiency* (25) for a set of nodes. This quantity allows us to test the properties of the set of nodes characterized by mutations in the network.

The global efficiency of a network is defined as

$$E_{global} = \frac{1}{n(n-1)} \sum_{\substack{i,j \in N \\ i \neq j}} \frac{1}{d_{ij}}$$

where $d_{ij}$ is the geodesic distance (length of shortest directed path) from node $j$ to node $i$, $N$ is the set of all nodes in the network, and $n$ is the number of nodes in the network. Note that $d_{ij} = \infty$ if no directed path exists from $j$ to $i$, and that $0 \leq E_{global} \leq 1$. The inverse of the global efficiency gives a measure of the average geodesic distance between nodes in the network, with the average giving more weight to small distances. Unlike the characteristic path length, the global efficiency is finite for disconnected networks and weakly connected directed networks.

We define the *intraset efficiency* of a set of nodes $M$ as

$$E_M = \frac{1}{m(m-1)} \sum_{\substack{i,j \in M \\ i \neq j}} \frac{1}{d_{ij}}$$

where $m$ is the number of nodes in $M$, and paths from $j$ to $i$ are allowed to pass through any nodes in $N$, including those not in $M$. This measures how efficiently nodes within $M$ communicate with each other. Because the sum is performed over $M$ only, it is a set property rather than an average of single node properties. $E_M$ is normalized so that $0 \leq E_M \leq 1$. $E_M = 0$ occurs if and only if $d_{ij} = \infty$ for all nodes $i$ and $j$ in $M$, and $E_M = 1$ occurs if and only if $M$ is a clique (all nodes are connected to all other nodes by a bidirectional edge). $E_M$ thus measures how tightly connected $M$ is. Unlike the clustering coefficient, however, the intraset efficiency includes contributions from nodes which are separated by a distance of more than one unit.

Defining $M$ to be the set of 21 genes commonly mutated in AML, the intraset efficiency was computed for $M$ as well as for a control set of 10 million randomly generated sets of 21 genes. $E_M$ was found to be $0.2979$, $1.8\%$ larger than the largest intraset efficiency of the random sets observed and $145\%$ larger than the mean. A skew normal probability distribution was fitted to a histogram of the randomized sets with $R^2 = 0.999982$, and an approximate right-tailed p-value of $7.3 \times 10^{-8}$ was obtained for $E_M$. A more conservative estimate of the p-value was obtained by assigning a random direction to each protein-protein interaction whose true direction is unknown (rather than using an undirected edge). Adding edges to a network guarantees an increase in the global efficiency as well as the average intraset efficiency. The genes in $M$ communicate with their neighbors predominantly through protein-protein interactions, and some of the PPI edges are listed as undirected in the AML 2.1 network because their true directions are unknown. To ensure that these undirected edges were not the sole cause of the statistical significance of $E_M$, a new network was constructed in which each undirected PPI edge was assigned a random direction. The significance of $E_M$ was then calculated from the new network using 10 million random sets of 21 nodes as a control. Only 63 of these sets had intraset efficiencies greater than $E_M$, which gives an estimated p-value of $63/10^7 = 6.3 \times 10^{-6}$. Calculating the p-value using this method does not assume a specific form for the distribution of intraset efficiencies.

## Response of primary AML cells to a library of kinase inhibitors

The viability response to kinase inhibitors of primary AML samples obtained from two clinical centers (Scripps Health and UCSD Moores Cancer Center) was studied varying several steps of the experimental approach. This has the advantage of not limiting the relevance of our findings to only one set of experimental conditions.

Leukemia cells were obtained from patients with newly diagnosed or relapsed AML under IRB-approved protocols and with informed consent. Mononuclear cells from the samples were isolated by centrifugation through Ficoll-Paque™ PLUS (17-1440-02, GE Healthcare) exactly as per manufacturer's instructions. For both sets of clinical samples the same EMD kinase inhibitor library (EMD Millipore), composed of 244 inhibitors, was screened for its cytotoxic effects on the AML patient cells.

The isolated patient cells were plated at a density of 8000 or 25000 cells/well in 384-well plates (Greiner Bio-One) in RPMI-1640 (Hyclone) supplemented with 10% fetal bovine serum (FBS, Hyclone), 100 U ml$^{-1}$ penicillin, and 100 µg ml$^{-1}$ streptomycin or in mTeSR1 (STEMCELL Technologies). Subsequently, the cells were treated with the 244 library compounds at 1.25 microM or 10 microM final concentrations, and cultured in a humidified 5% $CO_2$ atmosphere for 72 or 96 hours. At the end of the culture, cell viability was measured either using the ATPlite assay (PerkinElmer) or using the equivalent CellTiterGlo (Promega) according to the manufacturers' protocols. Luminescence was read using the Analyst HP (Molecular Devices) or BioTek Synergy2 plate readers.

## Regression analysis with KIEN

We have recently developed a method (35) that integrates information contained in drug-kinase profiling with *in vitro* screening. The method uses the *in vitro* cell response of single drugs as a training set to build linear and nonlinear regression models. For each kinase, the regression provides a coefficient score $\beta_k$ measuring the sensitivity of cells to alterations in the activity of that kinase. We have explored the correlation between this $\beta_k$ coefficient and three measures of centrality (degree, betweenness centrality and PageRank) for the same kinases according to the AML 2.1 network topology. The calculation of the $\beta_k$ coefficients used the drug response screening from a library of 141 kinase inhibitors measured on primary cells from eleven AML patients and drug kinase profiling data from a published database (36). The *in-vitro* drug response data from the eleven patients were averaged in the analysis. We calculated Pearson Correlation and Spearman Rank Correlation on a set of 101 kinases with positive $\beta_k$ present both in AML 2.1 and in the profiling dataset. The $\beta_k$ coefficients obtained from this analysis are given in supplementary Table 9S.

Supplementary data and software are available at www.leukemianetworks.org

# RESULTS

## Methodological Results

The details of the datasets we used and the corresponding numbers of genes are shown in Table 1. Table 1S also shows the number of interactions included in AML 2.1 obtained from each of the five gene expression datasets. As expected from the more quantitative nature of RNA-seq, the datasets obtained with this technique were more informative, providing more reproducible interactions compared to the three microarray datasets, even when the number of samples was comparable or lower.

Table 2 shows that the optimized CV cutoff increased the number of validated TRANSFAC interactions (TI hits) identified in almost all cases and specifically in every case where Pearson Correlation was used. Interactions were ordered according to the measurements provided by each method, and the significance of the CV cutoff was tested using the two-tailed student t-test of the top 100 and top 1000 interactions, before and after CV cutoff. The CV cutoff in both top 100 and top 1000 resulted in significant increases in TI hits with p-value < 0.0001.

The run times of our optimized Pearson Correlation method and of three previously published network reconstruction methods, ARACNE (18, 19), TIGRESS (20) and GENIE3 (21), were estimated using the same hardware and datasets and are shown in Figure 2S. The comparison shows a speed advantage of several orders of magnitude for the optimized Pearson Correlation method we present here. Adding another method after optimized correlation identifies 11-15 % more interactions (Table 3). As shown in Table 2S, however, these interactions are not the same for every method added. Figure 4S shows an example of a non-linear relationship that has been identified as a network interaction by the three additional methods listed in Table 3, but not by our method. As indicated in Table 3 these methods can miss an even larger number (14-60 %) of validated interactions found by optimized correlation.

**Reproducibility Results (Overlap)**

Table 4 shows that the reproducibility of our method (measured by the number of interactions found in more than one expression dataset) is much higher than that of randomly generated TFG and PPI subnetworks. The probability of finding the number of interactions reported 2 or more times is much lower than 0.01 for both the TFG and the PPI subnetworks. Random simulations and the exact method described in the "overlap analysis" section of the Methods provide similar estimates. None of the interactions found in only one gene expression dataset are included in the AML 2.1 network.

In addition to the reproducibility (overlap) of interactions, the TI ratio, defined as the number of TIs/number of interactions, was also examined, as shown in Table 3S for the TFG subnetwork. This ratio was increased compared to the initial dataset (before the application of the optimized correlation selection) even for interactions present in only one dataset and increased progressively for those appearing in two or more datasets.

The full lists of overlapping interactions for the TFG and PPI subnetworks are shown in Tables 4S and 5S. These tables also show that the average correlation is generally higher when the reproducibility increases (we show separately interactions found in 1, 2, 3, 4 or 5 datasets) but the distributions of these correlation values are not sufficient to separate the groups. In conjunction with the data already referred to in the Method section, these findings are consistent with the rationale we have used in the optimization of the network.

We also analyzed 7 more AML microarray datasets (GSE15434, 24006, 33223, 34860, 21261, 6891, 22845), which will be included in future version of the network, and measured the reproducibility of the interactions using all 12 datasets. The data are shown in Table 10S.

Using the 12 datasets we also measured the TI ratio for the first 10 overlap groups (the number of interactions was too small in groups 11 and 12). The data are presented in Supplementary Table 12S and show a monotonic (exponential) increase of the TI ratio with the number of overlaps, and therefore with the reproducibility of the interactions. We found a significant Spearman's correlation between the group number and the TI ratio with p<0.0001.

Remarkably, a similar increase as a function of the group number is also obtained when we measure, as a ratio, the interactions that, for each of the 10 overlap groups obtained from an analysis of 10 datasets, were found again twice when two more datasets were added. This ratio is a measure of the probability of reproducibility, and therefore of validity for interactions in each group, and was found to increase monotonically from overlap group 1 to overlap group 10 (Spearman correlation had p <0.0001) (Supplementary Table 12S). This probability was found to be well approximated by a single-parameter sigmoid function of the form $(1+e^{z-x})^{-1}$ where x is the overlap group and the fitting parameter z=6.15609 ($R^2$=0.9986). See Fig. 4A.

Reproducibility can also be studied within the groups shown in Figure 4A. The reproducibility of interactions in the overlap 2 group, ordered by an average rank obtained from their correlation coefficient, declined monotonically. We measured the interactions in this overlap 2 group that were found again after adding all possible combinations of two more datasets. Each of the interactions originally found in 2 datasets could now therefore remain in overlap group 2 or could be found in overlap groups 3 or 4. Assigning a value of 1 to those that progressed one step to group 3 and a value of 2 those that progressed two steps to group 4 the average score for the top 6000 interactions was 0.268 while for the remaining interactions in the group it was only 0.068, a decrease of almost 4-fold. This difference was highly statistically significant, using both a parametric and a non-parametric test (p<0.0001 with the Mann Whitney test).

We also studied this behavior in additional overlap 2 groups for datasets 6 to 10 and identified an exponential fitting that can be used to predict the distribution of the probability of reproducibility after the addition of two more datasets, as a function of the rank of an interaction within a group (Supplementary data and Figure 4B). The probability distribution is well approximated by $P(R,k) = ae^{-R/Q(k)}$ where *R* is the rank of the interaction within the group, and *Q(k)* is a characteristic decay rank that depends on the number of datasets k. The factor *a* takes into account the probability distribution normalization. We found that the characteristic decay rank scales with the number of possible interactions contained in each group and is well approximated by the relation $Q(k) = 515.04\binom{k}{2}$ $(R^2 = 0.998)$. Figure 4B shows that the correlation-based ranking within group 2 contains less information and is less significant when more datasets are added, since the top ranking interactions in this group become less reproducible.

These models and analyses can assist the choice of which interactions to select when the number of datasets increases and could also be used to build weighted networks.

## Biological Results

### Properties, Visualization and Gene Ontology cluster analysis of AML 2.1

The full list of TFG and PPI interactions in AML 2.1 is shown in Table 6S. The global network properties of AML 2.1 are shown in Table 5 (25). The AML 2.1 network contains the TFG and PPI subnetworks and is partially directed. MCODE clustering analysis found a total of 101 clusters. The complete list of clusters is shown in Table 7S. Table 6A also shows the differences between normal human hematopoietic cells and AML patients for the top 13 clusters, with corresponding GO (Gene Ontology) functional terms, and the p-values for these differences. The Fisher's exact test and false discovery rate (FDR) were performed on the clusters and 4 clusters were found to be expressed with a p-value <0.1 in either Normal subjects or AML patients. Two clusters related to immune response and cell cycle were found to be highly expressed in AML patients. On the other hand, one cluster related to translation and biosynthetic process was found to be highly expressed in normal human hematopoietic cells. Figure 5 shows the AML 2.1 network with the top MCODE clusters. Figure 5 also shows several other functions that are relevant to the cells of origin of AML, for example "Leukocyte and Lymphocyte activation". Table 6B shows similar comparisons using Fisher's exact test with RECON2 (32) metabolic pathway clustering. Eight RECON2 pathways were found to be differentially expressed.

### Receptors

We also examined the number of interactions for specific functional classes, including cellular receptors. The two most connected receptors, with degree (number of connections) higher than 200 were VDR (vitamin D receptor) and RXRA (retinoid X receptor, alpha) (Table 8S). As we mention in the Discussion these are known to have important roles in AML cells. We have also analyzed the two human AML RNA-seq datasets we use in this study (12, 13) and found that the coefficient of variation of receptors expression between different patients is in both cases approximately double that of other genes (p<0.0001).

### AML Mutations

The network is significantly enriched for common known AML mutated genes. It contains 21 out of 26 significantly mutated AML genes (2) even though it is composed of only 5667 genes/proteins ($p = 2.3 \times 10^{-8}$ for the enrichment). This shows that the network reconstruction method enriches for functionally relevant genes. Figure 6 shows the 21 common AML mutations included in the network and their first neighbors. This sub-network is highly connected with a total of 5 clusters and with the largest cluster containing 16 AML mutated genes. Figure 7 shows the mutations and their first neighbors within the AML 2.1 network. Comparing Figures 5 and 7 shows that the mutations co-localize with functional clusters of known relevance to cancer, including "cell cycle" and "DNA replication".

To examine the statistical significance of these measures, random subnetworks were generated. Random subnetworks consisting of 21 random genes and their first neighbors were less connected than the mutation subnetwork. They had an average of 16.2 clusters and an average size of 3.5 genes from the group of 21 in the largest cluster (p<0.0001 compared to the mutation subnetwork). Other network measurements were also computed with the same random subnetwork simulation, as shown in Table 7 and Supplementary figure 7S. The 21 mutations also had significantly higher values of 3 network centrality measures: degree (p = 0.015), betweenness centrality (p = 0.02) and PageRank (p = 0.01) (25).

A similar conclusion, with stronger statistical significance, is obtained by examining the intraset efficiency for the 21 mutations and for random sets of 21 genes. The intraset efficiency was clearly higher for the set of 21 mutations. Figure 8 shows that a skew normal probability distribution was fitted to a histogram of the randomized sets with $R^2 = 0.99$, and an approximate right-tailed p-value of $7.3\times10^{-8}$ was obtained. This measure indicates that the paths among the mutations are much shorter than for control sets. In other words the mutations can more easily exchange information.

As shown by visual inspection and comparison of Figures 5 and 7 and by calculating the clustering coefficient (table 7 and Figure 7S) the mutations do not, however, form a tight cluster. That is, they do not interact mainly among themselves.

Table 8 shows a summary of the GO functional enrichment analysis of the mutation subnetwork, obtained using DAVID (37). The full analysis is shown in Table 8S. The mutation subnetwork is composed of 21 common AML mutations and of their first neighbors, for a total of 257 genes, but a very similar list of GO terms is obtained by analyzing the first neighbors only (Table 8S), showing that the functional information is contained in the network and not only in the mutations. These functions are those commonly associated with cancer mutations, including DNA replication, cell cycle and cell death.

**Experimental validation using kinase inhibitors and AML primary samples**

Centrality measures can be used to rank kinases in AML 2.1. These results were compared to the response of AML primary cells to a library of 244 kinase inhibitors. A method we have recently developed, based on elastic net regression applied to kinase inhibitors, the KIEN method (35), was used to identify and rank according to a score $\beta_k$ (see Methods), the kinases responsible for the effects of the kinase inhibitors in primary AML cells from eleven patients.

We then calculated Pearson Correlation and Spearman Rank Correlation on a set of 101 kinases present both in AML 2.1 and in the drug response dataset (see Methods). Table 9A shows the correlation coefficients and significance values of betweenness centrality, degree, and PageRank with the KIEN parameter $\beta_k$, using Pearson and Table 9B shows the same three correlations using Spearman Rank Correlation. Betweeness centrality is significantly correlated with $\beta_k$ according to both methods while degree and PageRank were significant only with Pearson Correlation. A plot of the $\beta_k$ versus the three centrality measures for the 101 kinases used in this analysis is provided in the Supplementary Figures.

The top 10 kinases identified by the combined use of betweenness centrality, and PageRank with KIEN are shown in Tables 10 and 11. The most remarkable finding is the presence of a group of four kinases, CDK1, CDK2, CDK4 and CDK6 at the top of the independent analyses based on AML 2.1 centrality measures and on experimental data analyzed by KIEN (with significance of $p < 10^{-7}$ for betweenness centrality, PageRank and also degree, the details of the degree analysis are shown in the Supplementary data). The other kinases shown in Tables 10 and 11 are strong candidate targets for further experimental studies.

Specific literature support for the involvement of these targets in AML is analyzed in more depth in the Discussion but and additional level of statistical confirmation of our approach is obtained by showing that the number of relevant citations for each of the 101 kinase targets mentioned above in this section (obtained by searching Pubmed for the gene name and the term AML) is significantly correlated (using Spearman) with the combined average rank of the kinases obtained as shown in Tables 10 and 11. The p value is lower than 0.0003 for ranks obtained from all three centrality measures (betweenness centrality, degree, and pagerank).

## DISCUSSION

We have developed a fast, reproducible and scalable network reconstruction method, which is able to integrate biological datasets of different types. In the AML 2.1 network version we present here, both microarray and RNA-seq gene expression data and protein-protein interaction data were included.

Only interactions derived from at least two independent clinical datasets were selected for the network and some of the interactions found twice underwent further filtering. This is the only strategy that can correct for all possible types of noise, including biological, clinical and experimental variation. As can be seen from Table 4, this led to pruning of a large number of interactions, and, most likely, to a higher quality AML network. The alternative approach of pooling all the data and performing a single analysis would be much less tractable computationally, would be less efficient when a new dataset is added, would pose severe problems of normalization among studies and would be more prone to artifacts, because a small number of data points can greatly affect the correlation coefficient. Even in fields as diverse as particle physics (38, 39) and clinical drug development (40-43) performing multiple studies is considered a source of stronger evidence compared to pooling all resources in a single giant study.

The analysis of reproducibility in different datasets, which we call “overlap analysis”, can also provide a quantitative estimate of the probability of an interaction, based on the number of datasets in which it has been found. Figure 4B shows that within a group with the same reproducibility measure (that is containing interactions found in the same number of datasets) the value of the correlation coefficient for an interaction could predict the probability of being identified again when additional datasets were analyzed. As shown in Figure 4A, the same was true when comparing different groups. It is therefore clear that determining which interactions to include in the network is a trade-off between maximizing the confidence in the included interactions and building a network too sparse to have sufficient statistical power for meaningful analysis. The use of a weighted network is an alternative strategy that might allow the appropriate inclusion of a larger number of links (44).

The clustering analysis identified gene functions that are consistent with the tissue of origin of AML. The differences of AML vs normal cells gene expression for these clusters was also in the expected direction, for example glycolysis is well-known to be up-regulated in cancer cells.

Among the findings supporting the biological relevance of AML 2.1 are the observations that the network is significantly enriched for common known AML driver mutated genes (2) and that the mutation subnetwork is enriched for important cancer-related functions. Most notable among these are cell cycle related genes, presently one of the most active fields of drug development in many cancer types (45, 46).

The network properties of AML mutations we report are potentially useful for the understanding and therapy of cancer. It seems that mutated cancer genes not only are related to the functional categories we know well (47) but also have network properties of efficient communication among the set and of centrality, therefore being able to influence many other cell functions. They do not form a close cluster, where the genes preferentially interact only among themselves. The centrality findings we obtained are consistent with previous reports of the relevance of PageRank network measures to the identification of cancer biomarkers (48).

The targets shown in Tables 10 and 11 were identified using both the AML 2.1 network properties and the KIEN analysis of experimental drug response data from AML primary cells. The four targets with higher statistical significance were CDK1, CDK2, CDK4 and CDK6. CDK 4/6 inhibitors have been shown to be effective in phase II cancer clinical trials, some of which were presented at the ASCO and AACR 2014 meetings (49). One of these CDK 4/6 inhibitors, palbociclib, has received the “breakthrough therapy” designation by the FDA (50), which is intended to lead to accelerated approval. Several papers have also shown that CDK inhibitors are effective in AML cells (51-54).

The other targets shown in Tables 10 and 11 have also all been previously linked to AML and, in some cases, to cell cycle genes. LCK and LYN are part of the SRC family and CSK is a kinase acting on SRC. SRC family kinases have been implicated in AML by several authors (55, 56) and are targets of Dasatinib, which has been shown to be active on AML cells (57). LCK is also known to interact with and being phosphorylated by CDK1 (58). TYRO3 expression has been associated with AML (59) and the expression of his ligand identifies high-risk AML patients (60). CHEK1 is another important cell cycle gene and suggested target for cancer therapy (61), which has been shown to sensitize AML cells to cytarabine action in an RNAi screen (62). CHUCK (also known as IKK-alpha) is part of the cell cycle regulatory network together with CHEK1 (63) and also contributes to the regulation of cell death in AML cells (64). RPS6KA1 has been suggested as one of the mediators of the anti-apoptotic action of FLT3, one of the main AML mutations (65). Finally MAP2K2 (also known as MEK2) has a very important role in regulating CDK4/6 activity (66) and is often activated in AML cells (67).

The potential of the combined use of AML 2.1 analysis and KIEN is not simply to provide a list of a few targets to be completely inhibited. We can actually identify the optimal amount of inhibition of each target, which corresponds to the coefficients of the KIEN regression equation, for a large number of kinases. This can potentially lead to the type of precise and robust distributed control that is common in biology (for example by transcription factors or microRNAs (22)) but until now not in pharmacology. The kinase response in vitro of primary cells is however in part influenced by the culture conditions, which differ from the in vivo microenvironment (68) and obtaining additional independent confirmation using the AML 2.1 network properties is extremely useful.

This combined approach can also be used for personalized therapy. We show that useful data using hundreds of kinase inhibitors can be obtained using primary cells, and even more precise individual targeting information could be obtained using the larger libraries (composed of up to thousands profiled kinases (22)) that several pharma companies have at their disposal. This would represent a dynamic molecular profiling of leukemic cell response, potentially much more valuable than the static snapshot of present omics techniques. The network could also be personalized further, for example by using individual gene expression data to prune not significantly expressed gene and by giving a greater weight to mutations form a single patient and to their first neighbors within the network. An optimal kinase inhibitor combination could therefore be designed computationally (35), even in cases when the mutations would not be actionable, and then verified further by appropriate systematic testing using patient's cells (69, 70).

While our increasing appreciation of the heterogeneity of cancer mutations (71), both between and within patients, is a cause of concern for the development of generally effective therapies, the identification of their shared pattern of connections raises the hope that sufficiently large and precisely calibrated combinatorial therapies designed along the principles we have discussed might benefit a wide range of patients.

The network could also be used to identify the receptors likely to have the largest effects on AML cell viability. The most connected receptors include some with well-known effects in AML cells, supporting the relevance of the network model. Among these are several interleukin receptors and the interferon gamma receptor. The top two receptors for connectivity (degree) and other network properties are VDR and RXRA. The ligands for these receptors, vitamin D3 and retinoic acid, have in fact well-known effects on AML cell proliferation and differentiation (72, 73).

As we mentioned in the Introduction, we suggest that network reconstruction should also be socially scalable, in the sense of facilitating the integration of information from scientists of different backgrounds. This would be made easier by the adoption of the open source model for the continuous improvement of the networks and of the related software. Open source software is written by many (up to thousands) volunteer computer programmers publicly sharing and reviewing their work in real time as part of a self-organized community (74). Several fields of software development have seen the emergence of very successful open source approaches (74), for example the operating system Linux and the web server application Apache (74).

We report data comparing the method used for AML 2.1 with other common methods for network reconstruction (18-21). It is reasonable to conclude that since all methods share most validated interactions found in our test they might considered to be roughly equivalent, and it is certainly possible that combining multiple methods might be useful (15, 75). We would need to understand more about the biological significance of the interactions that are uniquely found by each method to do a more precise comparison. It is also possible that after adding more data all methods will eventually find essentially the same set of interactions. It is clear however from the run-time analysis (see also Supplementary data) and from considering the computational steps each method performs that the method described here is much faster. It has also been designed to be especially scalable, because most calculations do not need to be repeated when a new dataset is added. In addition, the portion of the method based on reproducibility in multiple datasets (the "overlap analysis") is also applicable to other network reconstruction strategies.

It has been stated by leaders in artificial intelligence and data mining that "invariably, simple models and a lot of data trump more elaborate models based on less data" (76, 77). Thus a case might be made for considering as our top priority the analysis of all existing gene expression datasets with the fastest and most scalable method that gives a reasonable performance in network reconstruction. We have shown that very useful information can be obtained in a study using only five datasets and it seems that we are not doing all we can for cancer patients if we leave existing data unutilized.

Planned future additions to the network include the use of HIPPIE (14), with the same optimization role for PPI as that played by TRANSFAC (17) for TFG (expected in version 3) and the addition of microRNA/target interactions and of metabolic networks (32) (expected in version 4). We also intend to use more AML datasets (potentially all published ones) and to explore subtypes of this acute leukemia, including pediatric AML. We then plan to extend the approach to other leukemias and eventually to other cancers and to other diseases. It will also be important to develop network models for normal cell types to assist the design of selective therapies with reduced toxicity. This will allow the development of comparative network analysis. For example, the evaluation of the general relevance of the network properties we describe for the AML mutations will only be possible when networks for many different cancer types will be reconstructed using comparable methods.

We therefore present a fast and scalable method for the reconstructions of intracellular networks that can contribute to the understanding of the network role of cancer mutations and to the identification of targets for therapeutic interventions, also in combination with complementary statistical analyses of experimental data.

## FIGURE LEGENDS

### Figure 1. Basic workflow to generate the AML network (version 2.1)

[a] : Only genes with expression level higher than 1.0 RPKM were selected and then log transformation was used to normalize the RNA-seq data.

[b] : The Threestep function with default setting from the affyPLM package by Bioconductor was used for microarray data processing. Additionally, multi-probe to gene mapping used median probe expression.

[c] : Four methods were compared: Pearson correlation; Aracne; TIGRESS; GENIE3. The methods were compared using run-time simulations and the TRANSFAC Interaction (TI) hit rate using test datasets.

[d] : The average for the Poisson distribution was 2 TIs per interval. Only interactions within intervals with p-value lower than 0.10 were selected for later analysis.

[e, f] : Only reproducible interactions were selected. A further selection was based on the probability of finding the same interaction multiple times only by random chance.

### Figure 2. CV Optimization.

Panels A and B show an example of a retained gene target (LCK) and of an eliminated gene target (GP5) of the same transcription factor (ETS1). GP5 was eliminated from the network because of the low coefficient of variation. It s clear that the transcription factor is not likely to increase the expression of the target. Panels C and D show an example, for one of the datasets (Eppert), of the two-steps optimization, with finer resolution in D. The coefficient of variation (CV) value is optimized to give the highest TI rate.

### Figure 3. TFG and PPI interactions ranked by correlation values.

Panel A. Poisson statistic used for the selection of TFG interactions. The Panel shows the TI hits used for the Poisson distribution selection in the case of the CV-optimized Eppert dataset. Bins are ranked by correlation values, decreasing from left to right. The red line indicates the cutoff. Only interactions with correlation values above (to the left of) the cutoff were selected. These correspond to bins with a higher number of TIs, which are the validated TRANSFAC interactions.

Panel B. PPI interactions and HIPPIE hits**.** The Panel shows the number of HIPPIE Interactions hits within 15,000,000 random interactions from the Eppert dataset.

Panel C. PPI interactions and HIPPIE hits. The Panel shows HIPPIE Interactions hits within the first bin (50,000 interactions) in A, with finer resolution. Bins are ranked by correlation values decreasing from left to right. The analysis is from a randomly selected subset corresponding to about 10% of all possible correlations of the Eppert dataset. As for the TFG subnetwork shown in Panel A, also for the PPI subnetwork bins corresponding to interactions with a higher correlation coefficient contain a higher number of validated interactions obtained from the HIPPIE database.

**Figure 4. Reproducibility analysis.**

Panel A: Reproducibility Probability in 10 groups. The figure shows the probability that an interaction found in overlap group x (horizontal axis) using 10 datasets is found in group x+2 when 12 datasets are used. This probability gives an estimate of reproducibility for group 1 to group 10. The 10 points were found to be well approximated by a single-parameter sigmoid function of the form 1/(1+Exp(z-x)) where x is the overlap group and the fitting parameter z=6.15609 ($R^2$=0.9986).

Panel B Reproducibility Probability Distribution within group 2. The figure shows the distribution of the probability of reproducibility (after the addition of two more datasets) as a function of the rank R of an interaction within group 2 with k=5 to 10 datasets. The fitting to the probability distribution is of the form P(R,k)=*a Exp[-R/Q(k)],* where *R* is the rank of the interaction within the group, and *Q(k)* is a characteristic decay rank that depends on the number of datasets k. The factor *a* takes into account the probability distribution normalization. This figure indicates that the correlation-based ranking within group 2 contains less information and is less significant when more datasets are added, since the top ranking interactions in this group become less reproducible.

**Figure 5. AML Network 2.1 with the 13 main clusters.**

AML 2.1 is shown with 13 functional clusters highlighted. The clusters had significant differences between AML patients and controls. See Table 8 for a detailed description of the functions associated with each cluster.

**Figure 6. Mutation subnetwork.**

This subnetwork is composed of 21 common AML mutations and their first neighbors. The red dots are the mutations, the blue dots are their first neighbors, the blue edges are TFG interactions and the green edges are PPI interactions.

**Figure 7. AML Network 2.1 with the mutation subnetwork.**

AML 2.1 is shown with the 21 common AML mutations and their first neighbors highlighted. The red dots are the mutations and the pink dots are their first neighbors. The mutation subnetwork overlaps the region where in figure 5 we see the clusters for cell cycle, translation and DNA replication.

**Figure 8. Intraset efficiency.**

The intraset efficiency of the 21 genes commonly mutated in AML cells as well as 10 million randomly generated sets of 21 nodes for a control. The vertical axis shows the probability density. The histogram was built using 50 bins of uniform width. The red curve is the right-skewed normal distribution fitted to the random data, which has $R^2 = 0.999982$. The mutation intraset efficiency is greater than the intraset efficiency of all random sets examined.

# FIGURES

Figure1

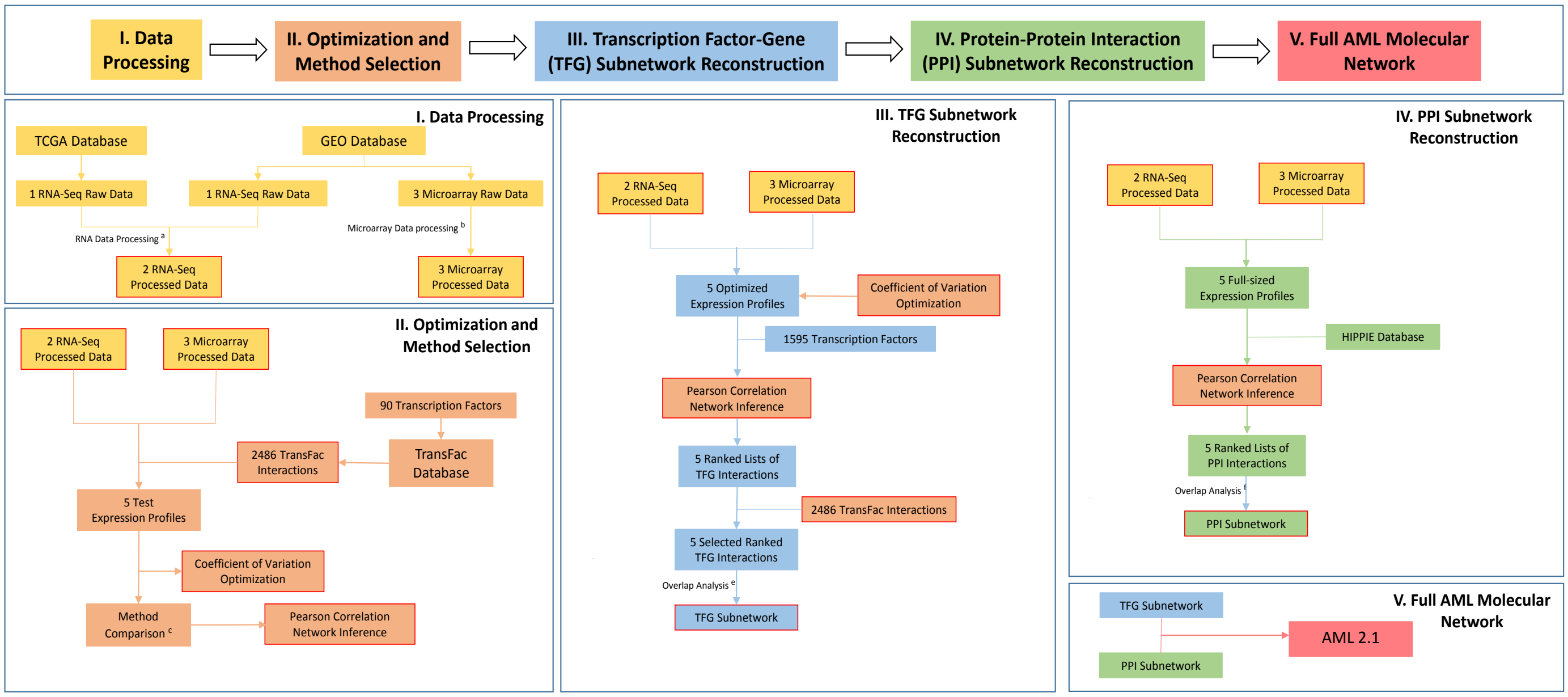

Figure 2

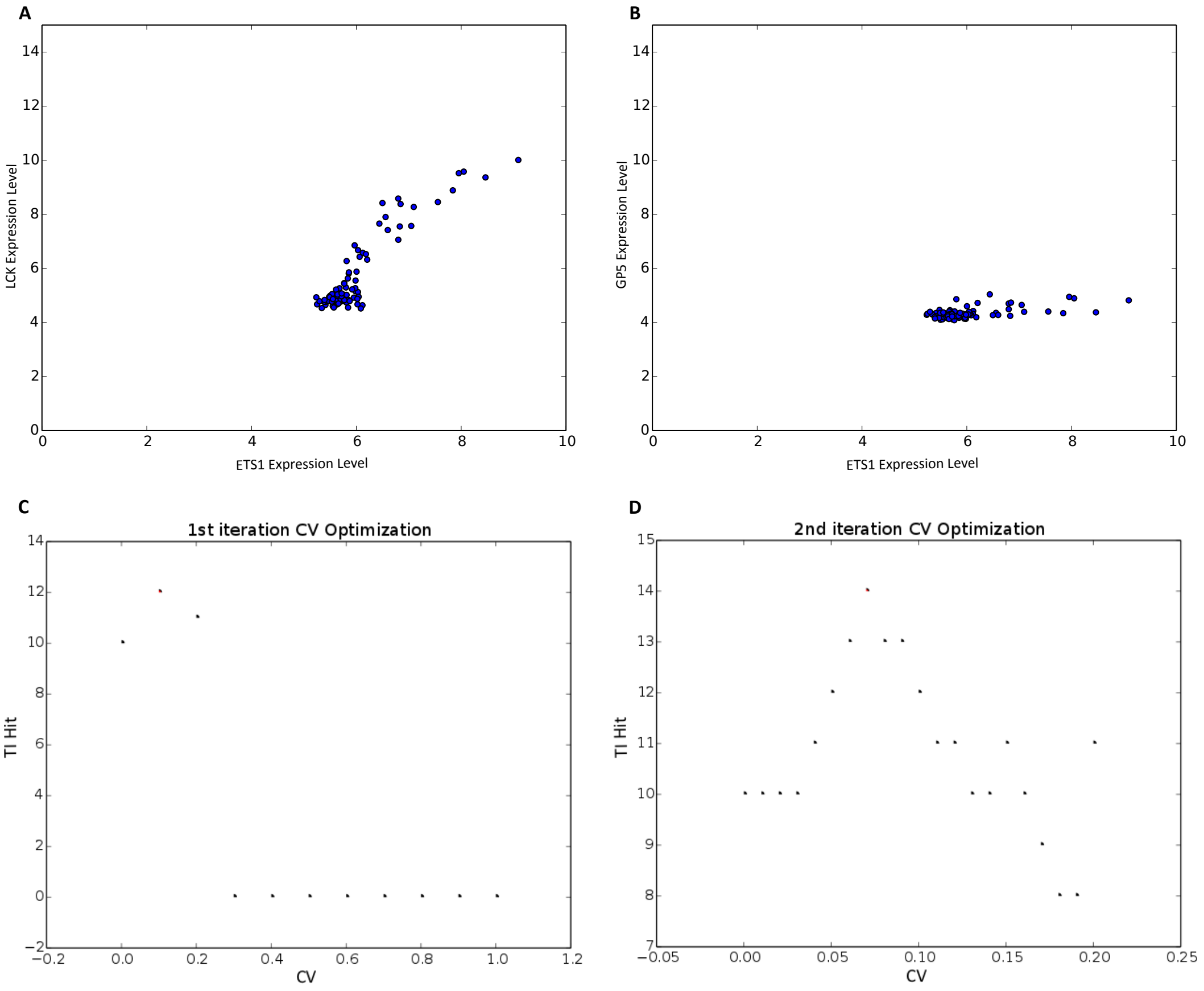

Figure 3

A

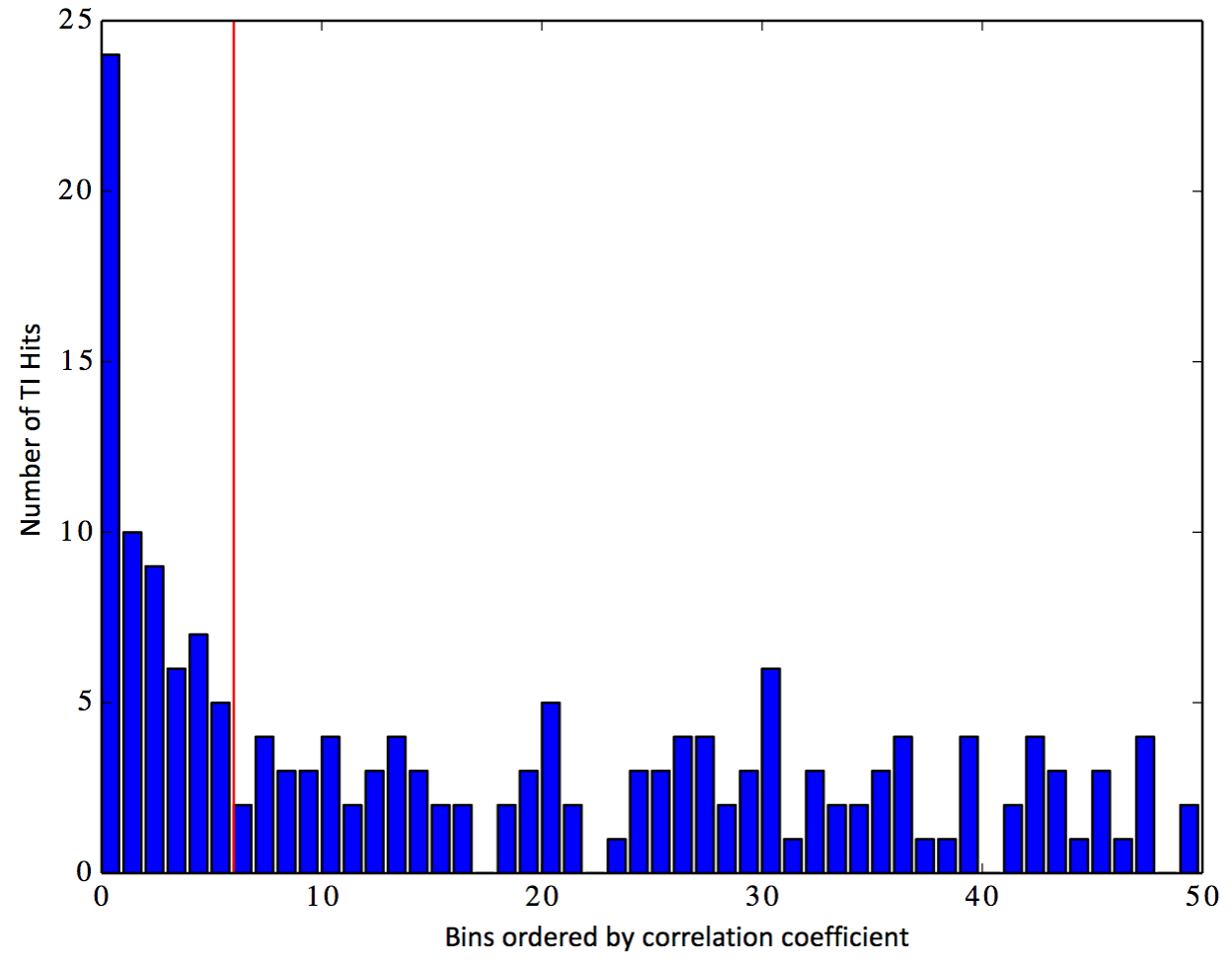

B

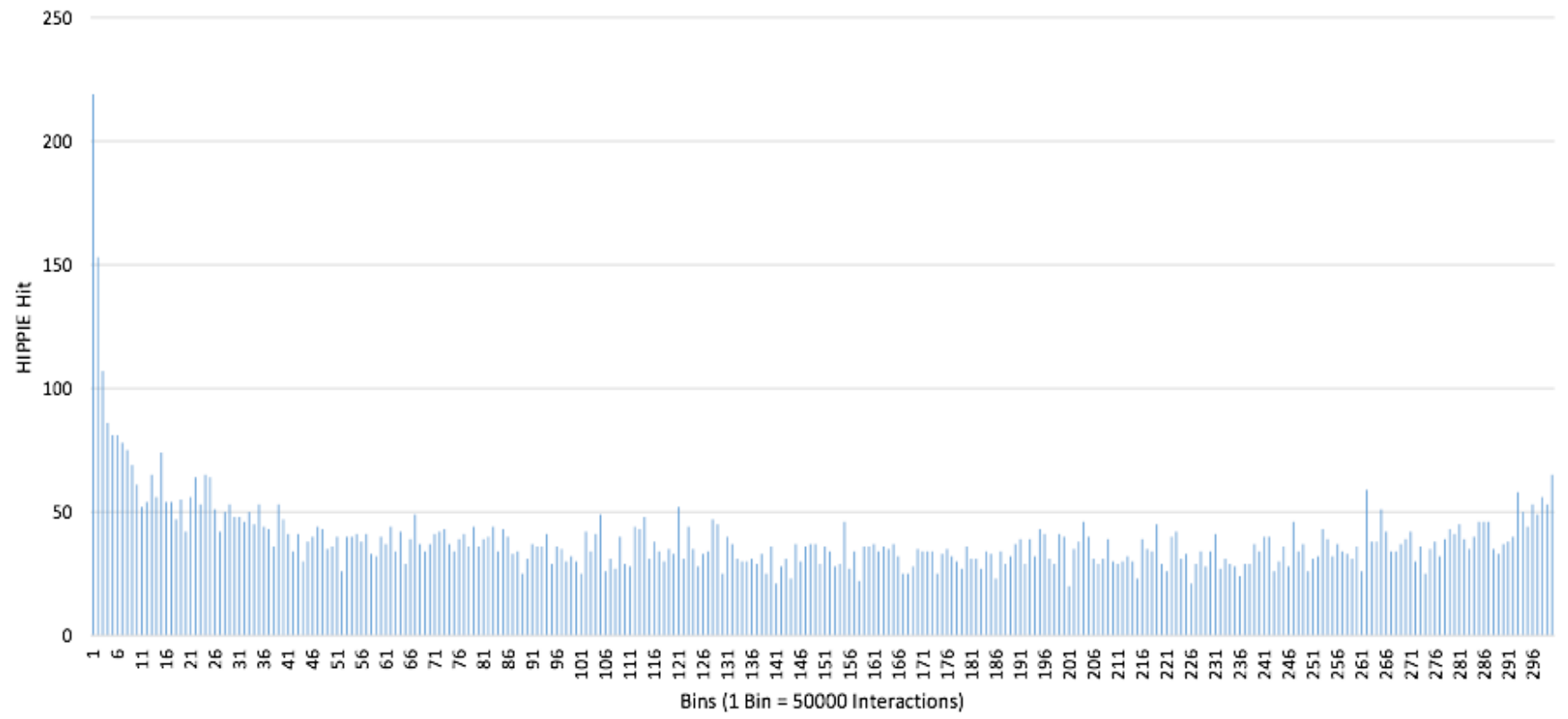

C

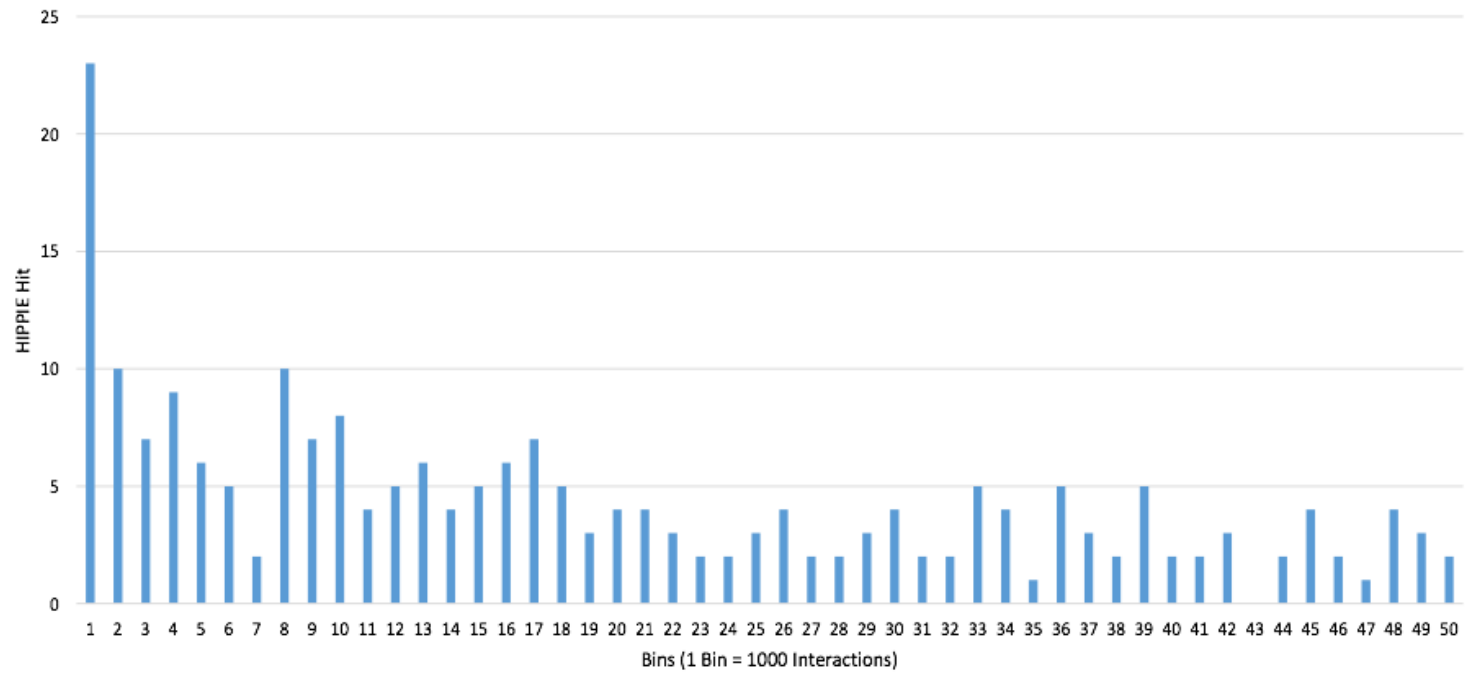

Figure 4

**A**

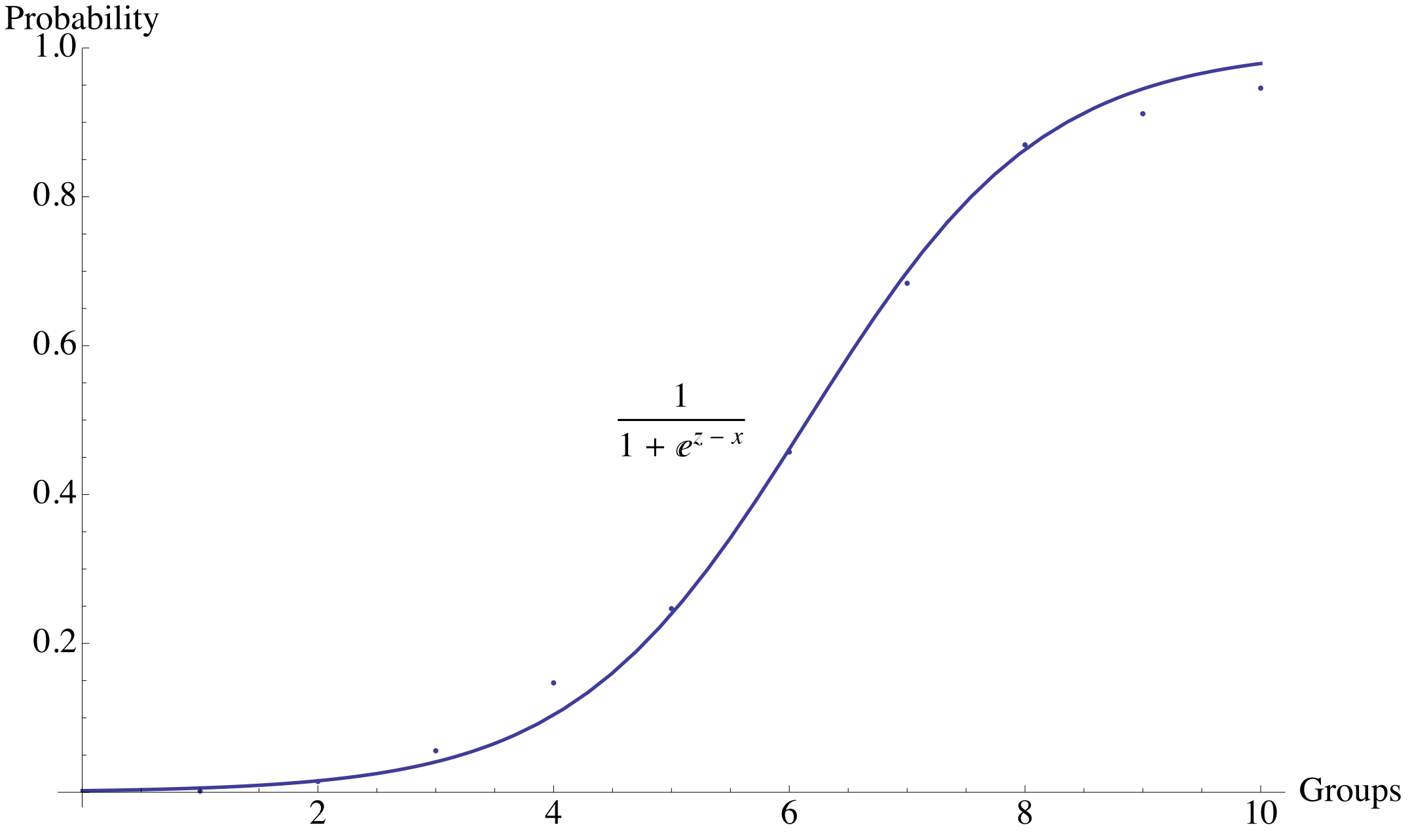

**B**

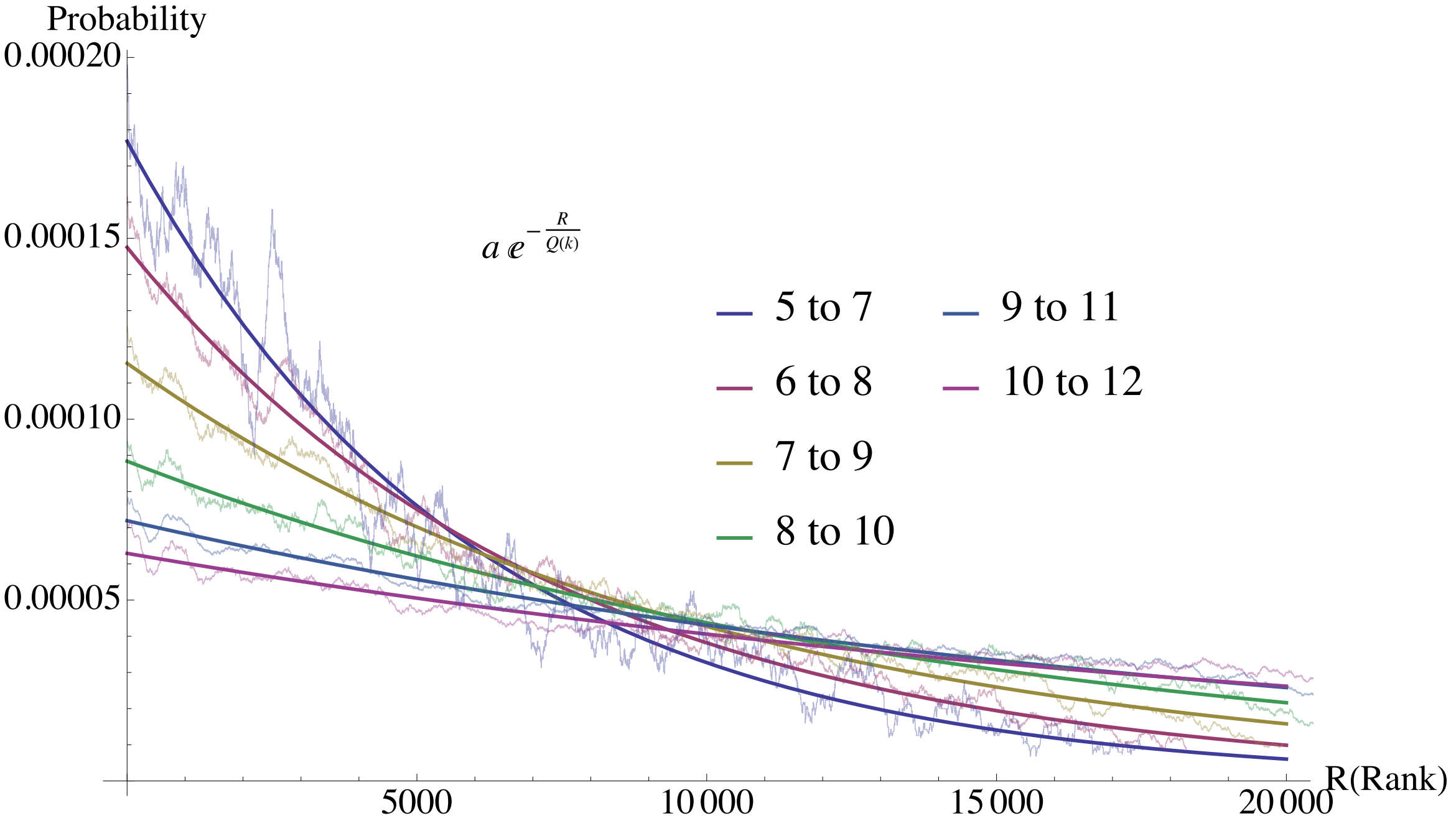
Probability
0.00020
0.00015
0.00010
0.00005
$a\, e^{-\frac{R}{Q(k)}}$
5 to 7
9 to 11
6 to 8
10 to 12
7 to 9
8 to 10
5000
10 000
15 000
20 000
R(Rank)

Figure 5

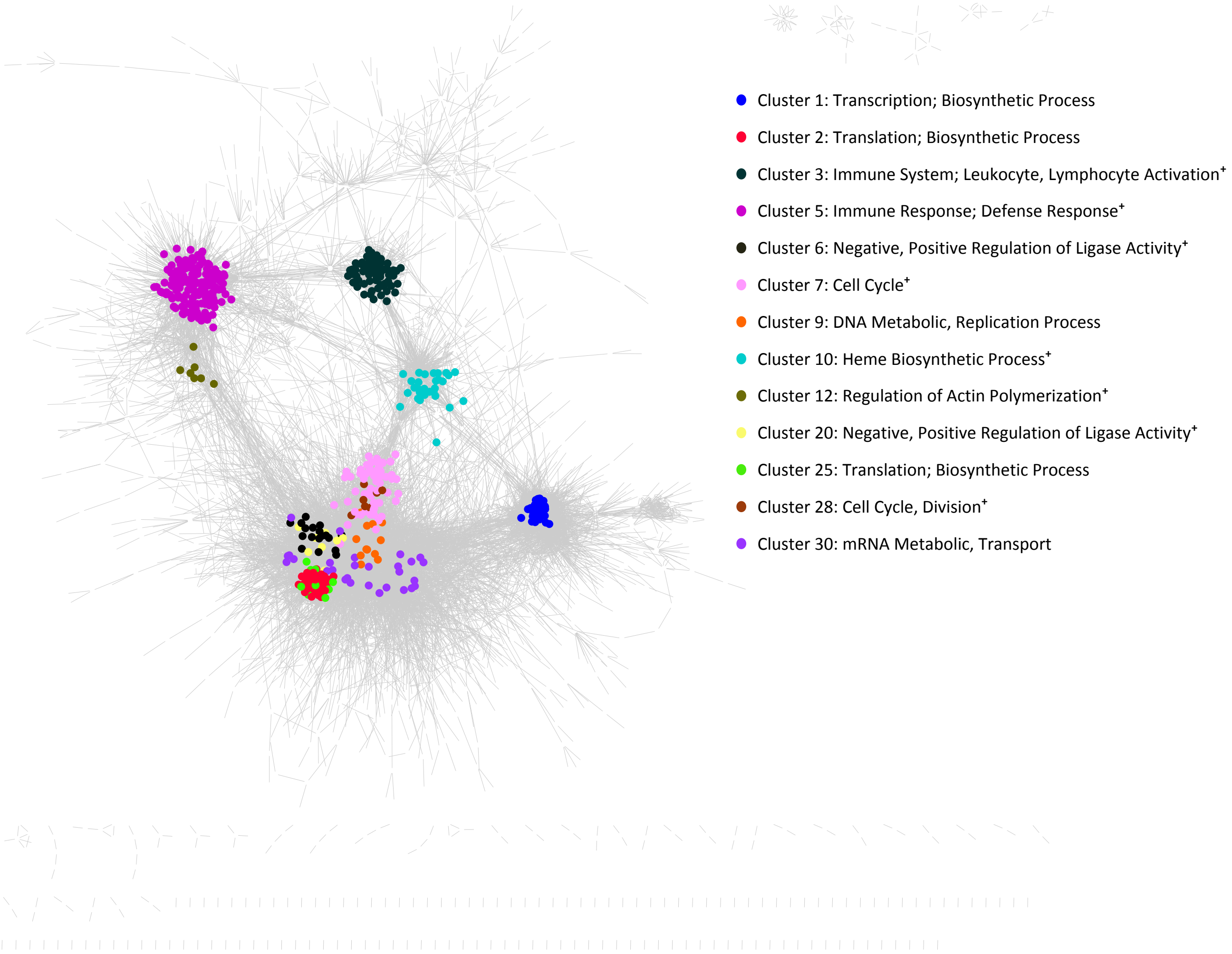

Cluster 1: Transcription; Biosynthetic Process
Cluster 2: Translation; Biosynthetic Process
Cluster 3: Immune System; Leukocyte, Lymphocyte Activation+
Cluster 5: Immune Response; Defense Response+
Cluster 6: Negative, Positive Regulation of Ligase Activity+
Cluster 7: Cell Cycle+
Cluster 9: DNA Metabolic, Replication Process
Cluster 10: Heme Biosynthetic Process+
Cluster 12: Regulation of Actin Polymerization+
Cluster 20: Negative, Positive Regulation of Ligase Activity+
Cluster 25: Translation; Biosynthetic Process
Cluster 28: Cell Cycle, Division+
Cluster 30: mRNA Metabolic, Transport

Figure 6

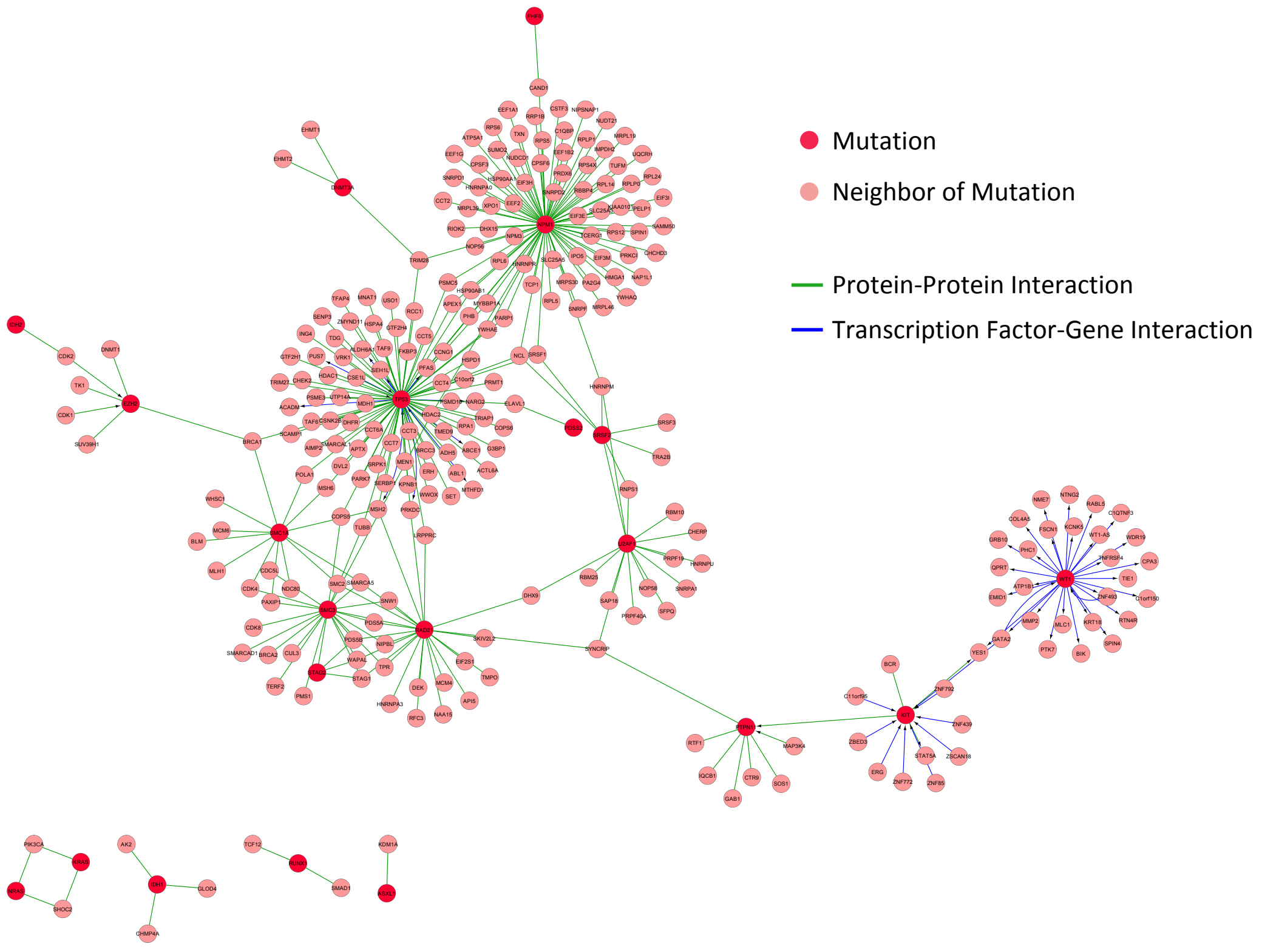
Mutation
Neighbor of Mutation
Protein-Protein Interaction
Transcription Factor-Gene Interaction

Figure 7

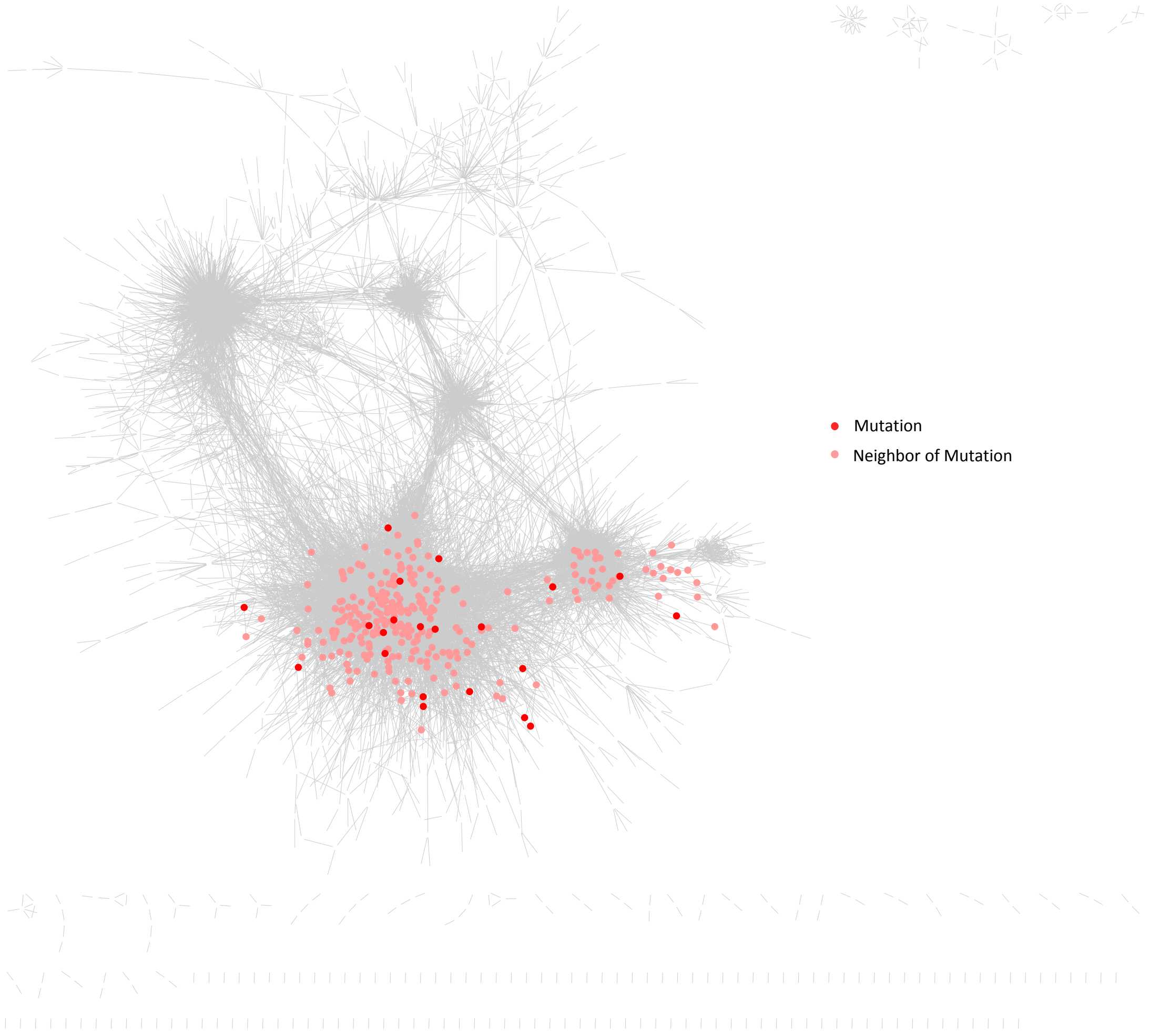

Figure 8

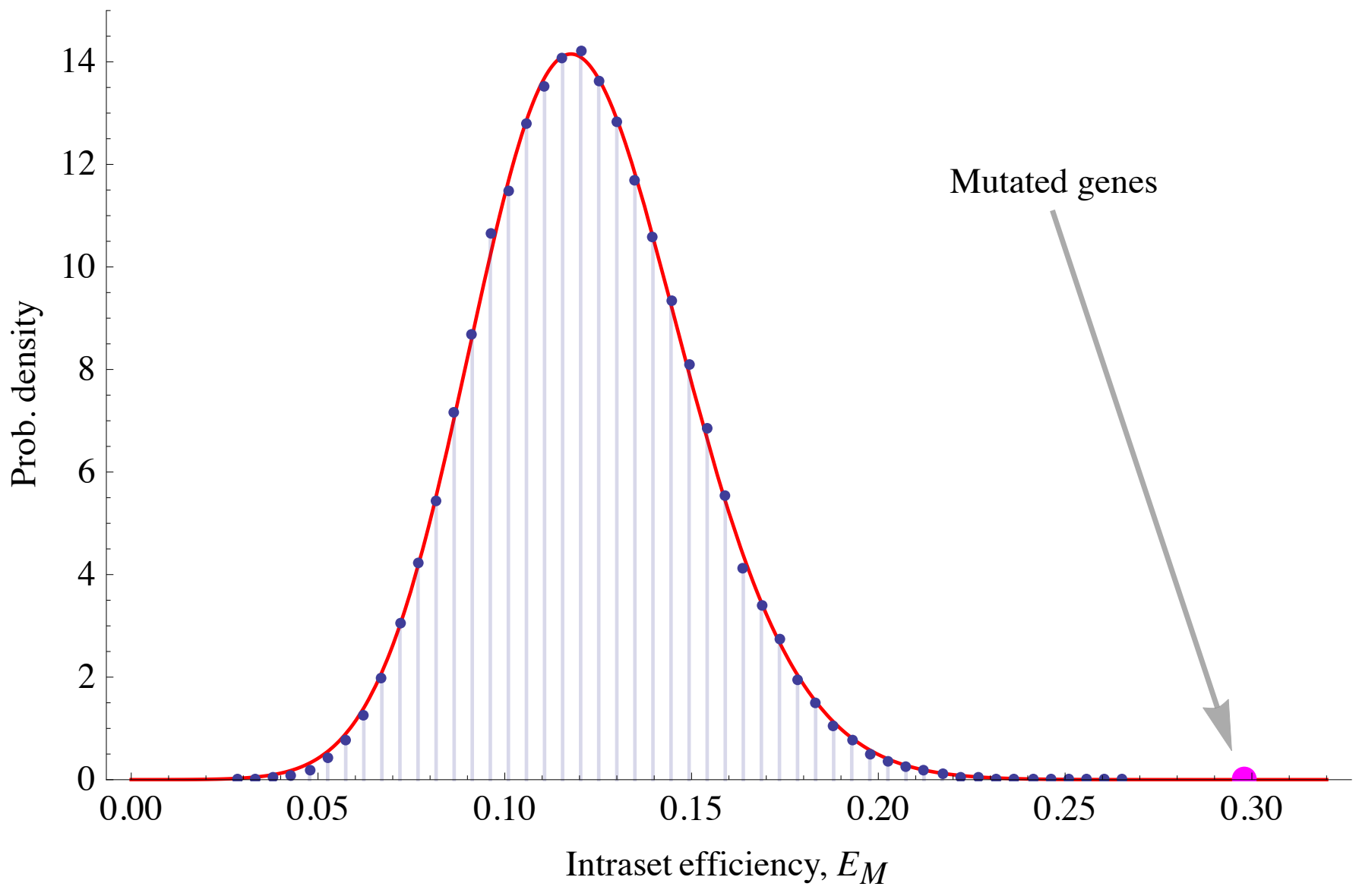

## TABLES

| Dataset | Technique | GSE ID | Number of samples | Full-Size Expression Profile. Number of genes | Test Expression Profile. Number of genes | CV Optimized Expression Profile. Number of genes |
|---|---|---|---|---|---|---|
| Eppert | Microarray | GSE30377 | 93 | 12495 | 1208 | 3663 |
| Metzeler | Microarray | GSE12417 | 163 | 12495 | 1208 | 4535 |
| Valk | Microarray | GSE1159 | 293 | 12496 | 1208 | 3388 |
| Macrae | RNA-seq | GSE49642 | 43 | 11737 | 785 | 3154 |
| TCGA | RNA-seq | NA | 179 | 12917 | 881 | 4332 |

**Table 1. Expression datasets used for the study, for both the TFG and the PPI subnetworks.** Three microarray and one RNA-seq datasets were downloaded from GEO. The TCGA RNA-seq LAML dataset was downloaded from the TCGA Data Portal.

| Network Inference Method | Data Source | Data Type | Top 100 TI Hit Pre CV | Top 100 TI Hit Post CV | Top 1000 TI Hit Pre CV | Top 1000 TI Hit Post CV |
|---|---|---|---|---|---|---|
| Pearson Correlation | Eppert | Microarray | 10 | 13 | 61 | 67 |
| Pearson Correlation | Macrae | RNA | 9 | 14 | 44 | 63 |
| Pearson Correlation | Metzeler | Microarray | 11 | 19 | 23 | 62 |
| Pearson Correlation | TCGA | RNA | 5 | 18 | 56 | 60 |
| Pearson Correlation | Valk | Microarray | 12 | 22 | 35 | 72 |
| Aracne | Eppert | Microarray | 6 | 8 | 45 | 43 |
| Aracne | Macrae | RNA | 7 | 7 | 28 | 35 |
| Aracne | Metzeler | Microarray | 3 | 14 | 30 | 47 |
| Aracne | TCGA | RNA | 9 | 11 | 56 | 46 |
| Aracne | Valk | Microarray | 11 | 17 | 50 | 61 |
| GENIE3 | Eppert | Microarray | 13 | 16 | 53 | 57 |
| GENIE3 | Macrae | RNA | 11 | 13 | 53 | 59 |
| GENIE3 | Metzeler | Microarray | 18 | 18 | 43 | 64 |
| GENIE3 | TCGA | RNA | 13 | 17 | 67 | 61 |
| GENIE3 | Valk | Microarray | 18 | 19 | 54 | 74 |
| TIGRESS | Eppert | Microarray | 9 | 13 | 43 | 60 |
| TIGRESS | Macrae | RNA | 5 | 8 | 52 | 55 |
| TIGRESS | Metzeler | Microarray | 16 | 17 | 38 | 56 |
| TIGRESS | TCGA | RNA | 14 | 15 | 58 | 54 |
| TIGRESS | Valk | Microarray | 14 | 19 | 42 | 70 |

**Table 2. Number of TI (TRANSFAC interactions) Hits for the top 100 and the top 1000 interactions.** Interactions were ordered by the values provided by the different inference methods, before and after the correlation coefficient (CV) cutoff. The data were obtained using the test datasets. The Table shows the increase after the CV cutoff.

| Method | TI Interactions shared with correlation | TI Interactions unique to correlation | TI Discovery unique to correlation - Ratio | TI Interaction unique to other method | TI Discovery unique to other method - Ratio |
|---|---|---|---|---|---|
| ARACNE | 49 | 35 | 0.60 | 9 | 0.11 |
| GENIE3 | 72 | 12 | 0.14 | 13 | 0.15 |
| TIGRESS | 65 | 19 | 0.24 | 13 | 0.15 |

**Table 3. Additional information provided by the indicated methods compared with optimized correlation.** The last column shows the ratio of newly identified TIs to those found with the correlation method. This column shows that adding an additional method increases the number of validated interactions (TIs) already found with optimized correlation only by 11-15 %. The third data column shows the same ratio when optimized correlation is added to one of the other three methods.

| Number of Datasets where an interaction is present | Avg. Interactions in Random Simulations | Interactions in Random Model | Interactions in TFG Subnetwork | Interactions included in AML 2.1 | Significance of number of reproducible interactions |
|---|---|---|---|---|---|
| 1 | 179574 | 179579 | 129943 | 0 | NA |
| 2 | 612 | 611 | 17817 | 6117 | $p < 10^{-10}$ |
| 3 | 1 | 1 | 2505 | 2505 | $p < 10^{-10}$ |
| 4 | 0 | 0 | 1183 | 1183 | $p < 10^{-10}$ |
| 5 | 0 | 0 | 596 | 596 | $p < 10^{-10}$ |

**Table 4A. The number of TFG subnetwork interactions that are found in 1 or more datasets are compared to those found in randomly generated subnetworks.** None of the interactions listed in the 1 Dataset row were included in AML2.1. Only part of the interactions found 2 times were included.

| Number of Datasets where an interaction is present | Avg. Interactions in Random Simulations | Interactions in Random Model | Interactions in PPI Subnetwork | Interactions included in AML 2.1 | Significance of number of reproducible interactions |
|---|---|---|---|---|---|
| 1 | 45836 | 45825 | 13754 | 0 | NA |
| 2 | 3 | 9 | 6794 | 6794 | $p < 10^{-10}$ |
| 3 | 0 | 0 | 2487 | 2487 | $p < 10^{-10}$ |
| 4 | 0 | 0 | 1705 | 1705 | $p < 10^{-10}$ |
| 5 | 0 | 0 | 844 | 844 | $p < 10^{-10}$ |

**Table 4B. The number of PPI subnetwork interactions that are found in 1 or more datasets are compared to those found in randomly generated subnetworks.** None of the interactions listed in the 1 Dataset row were included in AML2.1.

| | AML 2.1 |
|---|---|
| Nodes | 5667 |
| Edges | 22218 |
| Global efficiency | 0.1215 |
| Average clustering coefficient | 0.1983 |
| Transitivity | 0.2043 |
| Assortativity | -0.2008 |
| Betweenness centrality | 0.00054 |

**Table 5. Network Properties of AML 2.1.**

| Cluster ID | Representative GO Term | Higher Expression In | p-value |
|---|---|---|---|
| 5 | Immune Response; Defense Response | AML | 7.07E-18 |
| 2 | Translation; Biosynthetic Process | Normal | 8.46E-06 |
| 7 | Cell Cycle | AML | 0.00017 |
| 1 | Transcription; Biosynthetic Process | Normal | 0.012 |
| 3 | Immune System; Leukocyte, Lymphocyte Activation | AML | 0.031 |
| 20 | Negative, Positive Regulation of Ligase Activity | AML | 0.038 |
| 9 | DNA Metabolic, Replication Process | Normal | 0.038 |
| 25 | Translation; Biosynthetic Process | Normal | 0.039 |
| 10 | Heme Biosynthetic Process | Normal | 0.061 |
| 6 | Negative, Positive Regulation of Ligase Activity | AML | 0.062 |
| 12 | Regulation of Actin Polymerization | AML | 0.069 |
| 28 | Cell Cycle, Division | AML | 0.069 |
| 30 | mRNA Metabolic, Transport | Normal | 0.087 |

**Table 6A. Top 13 MCODE Clusters (P-value < 0.10).** The Fisher's exact test was used to compare the expression profile of AML and normal hematopoietic cells.

| RECON2 Pathways | Higher Expression In | P-value |
|---|---|---|
| Oxidative phosphorylation | AML | 0.0070 |
| Heme synthesis | Normal | 0.011 |
| Glycolysis/gluconeogenesis | AML | 0.016 |
| Transport, lysosomal | AML | 0.032 |
| N-glycan synthesis | Normal | 0.034 |
| NAD metabolism | AML | 0.038 |
| Selenoamino acid metabolism | Normal | 0.061 |
| Pentose phosphate pathway | AML | 0.065 |

**Table 6B. Top 8 RECON2 pathways (P-value < 0.10)**. The Fisher's exact test was used to compare the expression profile of AML and normal hematopoietic cells.

| Measurements | Mutations Mean | Control Mean | Control Median | Control STD | P-value |
|---|---|---|---|---|---|
| Clustering Coefficient | 0.130 | 0.198 | 0.193 | 0.0664 | 0.846 |
| Local Efficiency | 0.202 | 0.233 | 0.229 | 0.0723 | 0.648 |
| Degree | 27.952 | 11.799 | 10.619 | 5.338 | 0.015 |
| In-Degree | 13.476 | 5.894 | 5.476 | 2.169 | 0.0087 |
| Out-Degree | 14.476 | 5.905 | 4.952 | 3.753 | 0.0392 |
| Betweenness Centrality | 0.00206 | 0.0005 | 0.0004 | 0.0005 | 0.0207 |
| Eigen Centrality | 0.0027 | 0.0017 | 0.0003 | 0.0028 | 0.228 |
| PageRank | 10.942 | 5.313 | 5.0266 | 1.631 | 0.0125 |

**Table 7. Network measures of the 21 AML mutations set compared to controls (random gene sets of the same size).**

| GO Term | Description | Count | PValue | Fold Enrichment | FDR |
|---|---|---|---|---|---|
| GO:0006259 | DNA metabolic process | 43 | 4.28E-16 | 4.47 | 7.55E-13 |
| GO:0051276 | chromosome organization | 42 | 5.06E-16 | 4.56 | 9.44E-13 |
| GO:0006396 | RNA processing | 42 | 3.00E-14 | 4.04 | 5.08E-11 |
| GO:0006974 | response to DNA damage stimulus | 33 | 7.17E-13 | 4.66 | 1.22E-09 |
| GO:0007049 | cell cycle | 47 | 3.29E-12 | 3.19 | 5.58E-09 |
| GO:0033554 | cellular response to stress | 39 | 8.55E-12 | 3.63 | 1.45E-08 |
| GO:0006281 | DNA repair | 27 | 2.79E-11 | 5.00 | 4.73E-08 |
| GO:0016568 | chromatin modification | 23 | 1.16E-08 | 4.42 | 1.97E-05 |
| GO:0006260 | DNA replication | 18 | 1.20E-07 | 4.99 | 2.03E-04 |
| GO:0016570 | histone modification | 14 | 5.17E-07 | 6.04 | 8.77E-04 |
| GO:0010941 | regulation of cell death | 37 | 1.69E-06 | 2.39 | 0.0029 |
| GO:0034621 | cellular macromolecular complex subunit organization | 22 | 4.39E-06 | 3.24 | 0.0074 |
| GO:0045934 | negative regulation of nucleobase, nucleoside, nucleotide and nucleic acid metabolic process | 27 | 4.67E-06 | 2.78 | 0.0079 |

**Table 8. GO enrichment analysis for the mutation subnetwork of AML 2.1 (composed of 21 common AML mutations and their first neighbors, for a total of 257 genes).**

| Centrality Measures | Pearson Correlation Coefficient | p-value |
|---|---|---|
| Betweenness Centrality | 0.266 | 0.007 |
| Degree | 0.401 | 0.000032 |
| PageRank | 0.375 | 0.0001 |

**Table 9A. Pearson correlation of three centrality measures from AML 2.1 with experimentally obtained $\beta_k$ for 101 kinases.**

| Centrality Measures | Spearman Rank Correlation Coefficient | p-value |
|---|---|---|
| Betweenness Centrality | 0.234 | 0.018 |
| Degree | 0.178 | 0.075 |
| PageRank | 0.158 | 0.11 |

**Table 9B. Spearman rank correlation of three centrality measures from AML 2.1 with experimentally obtained $\beta_k$ for 101 kinases.**

| Kinase targets | Average rank | Betweenness Centrality rank | KIEN rank |
|---|---|---|---|
| **CDK2** | 2.5 | 1 | 4 |
| **CDK1** | 4 | 7 | 1 |
| **CDK4** | 6 | 4 | 8 |
| **CDK6** | 9 | 12 | 6 |
| LCK | 10.5 | 19 | 2 |
| LYN | 17.5 | 20 | 15 |
| CHEK1 | 18 | 15 | 21 |
| MAP2K2 | 18.5 | 24 | 13 |
| RPS6KA1 | 18.5 | 6 | 31 |
| CSK | 19 | 14 | 24 |

**Table 10. The top 10 kinase targets identified using the betweenness centrality measure from AML 2.1 and the KIEN analysis of experimental data.**

| Kinase targets | Average rank | PageRank rank | KIEN rank |
|---|---|---|---|
| **CDK1** | 2 | 3 | 1 |
| **CDK2** | 2.5 | 1 | 4 |
| **CDK4** | 5 | 2 | 8 |
| **CDK6** | 5.5 | 5 | 6 |
| TYRO3 | 11 | 13 | 9 |
| CHEK1 | 13.5 | 6 | 21 |
| LYN | 14.5 | 14 | 15 |
| CSK | 19.5 | 15 | 24 |
| RPS6KA1 | 23.5 | 16 | 31 |
| CHUK | 24 | 23 | 25 |

**Table 11. The top 10 kinase targets identified using the PageRank measure from AML 2.1 and the KIEN analysis of experimental data.**